\documentclass{article}
\usepackage{ijcai18}
\usepackage{times}
\usepackage{xcolor}
\usepackage{soul}
\usepackage[utf8]{inputenc}

\usepackage{amsmath}
\usepackage{amssymb}
\usepackage{algorithm}
\usepackage{algorithmicx,setspace}
\usepackage{algpseudocode}
\usepackage{graphicx}
\usepackage[subrefformat=parens,labelformat=parens]{subfig}
\title{Weighted Double Deep Multiagent Reinforcement Learning in Stochastic Cooperative Environments}

\author{
	Yan Zheng$^1$, 
	Jianye Hao$^1$, 
	Zongzhang Zhang$^2$, 
	\\ 
	$^1$ Tianjin University, Tianjin, China \\
	$^2$ Soochow University, Suzhou, China\\
	yanzheng@tju.edu.cn,
	jianye.hao@tju.edu.cn,
	zzzhang@suda.edu.cn
}

\begin{document}
	\maketitle
	\begin{abstract}
		Recently, multiagent deep reinforcement learning (DRL) has received increasingly wide attention. Existing multiagent DRL algorithms are inefficient when facing with the non-stationarity due to agents update their policies simultaneously in stochastic cooperative environments. This paper extends the recently proposed weighted double estimator to the multiagent domain and propose a multiagent DRL framework, named weighted double deep Q-network (WDDQN). By utilizing the weighted double estimator and the deep neural network, WDDQN can not only reduce the bias effectively but also be extended to scenarios with raw visual inputs. To achieve efficient cooperation in the multiagent domain, we introduce the lenient reward network and the scheduled replay strategy. Experiments show that the WDDQN outperforms the existing DRL and multiaent DRL algorithms, i.e., double DQN and lenient Q-learning, in terms of the average reward and the convergence rate in stochastic cooperative environments.
	\end{abstract}
	\section{Introduction}
	The goal of reinforcement learning (RL) is to learn an optimal behavior within an unknown dynamic environment, usually modeled as a Markov decision process (MDP), through trial and error \cite{sutton1998reinforcement}. Over the past years, deep RL (DRL) has achieved great successes. It has been practically shown to successfully master various complex problems \cite{mnih_playing_2013,mnih_human-level_2015}. To a large extent, these successes can be credited to the incorporation of the experience replay and target network that stabilizes the network training \cite{mnih2016asynchronous,mnih_playing_2013,mnih_human-level_2015,schaul_prioritized_2015,van2016deep,wang2015dueling}.
	
	Approaches like \cite{Bloembergen2011ETS,matignon2007,matignon2012independent,panait2006lenient,wei2016lenient} have been proposed by extending Q-learning to address the coordination problems in cooperative multiagent systems. They are able to achieve coordination in relatively simple cooperative multiagent system. However, none of them has been combined with deep learning techniques.
	
	Recently, increasing wide attention has been drawn in employing DRL in multiagent environments. Unfortunately, these multiagent DRL algorithms still suffer from two intrinsic difficulties in the interactive environments \cite{gupta2017cooperative,lanctot2017unified,matignon2012independent}: stochasticity due to the noisy reward signals; and non-stationarity due to the dynamicity of coexisting agents. The stochasticity introduces additional biases in estimation, while the non-stationarity harms the effectiveness of experience replay, which is crucial for stabilizing deep Q-networks. These two characteristics result in the lack of theoretical convergence guarantees of most multiagent DRL algorithms and amplify the difficulty of finding the optimal Nash equilibriums, especially in cooperative multiagent problems. 
	
	This work focuses on learning algorithms of independent learners (ILs) in cooperative multiagent systems. Here, we assume that agents are unable to observe other agents' actions and rewards \cite{claus1998dynamics}; they share a common reward function and learn to maximize the common expected discounted reward (a.k.a. return). To handle the stochastic and non-stationary challenges in the multiagent systems, we propose the weighted double deep Q-network (WDDQN) with two auxiliary mechanisms, the lenient reward network and the scheduled replay strategy, to help ILs in finding the optimal policy that maximizes the common return.
	
	Our contributions are three-fold. First, we extend weighted double Q-learning (WDQ) \cite{zhangweighted}, a state-of-the-art traditional RL method, to the multiagent DRL settings. Second, we introduce a lenient reward network inspired by the lenient Q-learning \cite{palmer2017lenient,panait2006lenient}. Third, we modify the exisitin prioritized experience replay strategy to stabilize and speed up the learning process in complex multiagent problems with raw visual inputs. Empirical results demonstrate that on a fully cooperative multiagent problem WDDQN with the new mechanisms indeed contribute to increasing the algorithm's convergence, decreasing the instability and helping ILs to find an optimal policy simultaneously.
	
	\section{Preliminaries}
	This section briefly introduces the definition of cooperative Markov games, Q-learning and its variants.
	
	\subsection{Cooperative Markov Game}
	Markov (stochastic) games, as an extension of repeated games and MDPs, provide a commonly used framework for modeling interactions among agents. They can be formalized as a tuple $<N,S,\mathbf{A},Tr,R_1,...R_N,\gamma>$. Here, $N$ is the number of players (or agents), $S$ is the set of states, $\mathbf{A}=A_1 \times ... \times A_N$ is the joint action set, where $A_i$ is the action space of player $i$, $Tr$ is the transition function $S\times \mathbf{A}\times S\rightarrow [0,1]$ such that $\exists s \in S, \exists a \in \mathbf{A}, \sum_{s^\prime \in S}Tr(s,\mathit{a}, s^\prime) = 1$, $R_i$ is the reward function $S\times \mathbf{A}\rightarrow \mathbb{R}$ for player $i$, and $\gamma \in \left[0, 1\right] $ is a discount factor. The state $s$ is assumed to be observable for all players. A fully cooperative Markov game is a specific type of Markov games where all agents receive the same reward under the same outcome, and thus share the same best-interest action.
	
	\subsection{Q-learning and Its Variants}
	\subsubsection{Q-learning} is based on the core idea of temporal difference (TD) learning \cite{sutton1988learning} and is well suited for solving sequential decision making problems \cite{claus1998dynamics,watkins1989learning}. Q-learning tries to find an accurate estimator of the Q-Values, i.e. $Q_t(s,a)$, for state-action pairs \cite{claus1998dynamics}. Each Q-value is an estimate of the discounted sum of future rewards that can be obtained at time $t$ through selecting action $a$ in state $s$. The iterative update formula is outlined in Equation \ref{eq:q-learning}:
	\begin{equation}\label{eq:q-learning}
	Q(s,a) \leftarrow Q(s,a) + \alpha [r + \gamma \max_{a^\prime}Q(s^\prime, a^\prime) - Q(s,a)],
	\end{equation}
	where $ r $ is the immediate reward and $\alpha \in [0,1)$ is the learning rate. The updating process always chooses the action $a^\prime$ with the maximum Q value and updates Q with the saved Q value. Once the process terminates, an optimal policy can be obtained by selecting the action with the maximum Q-value in each state \cite{bellman1958dynamic}. However, Q-learning uses a single estimator to estimate $E\{\max_{a^\prime}Q(s^\prime, a^\prime)\}$, which has been proved to be greater than or equal to $\max_{a^\prime}E\{Q(s^\prime, a^\prime)\}$ \cite{smith2006optimizer}. Thus, a positive bias always exists in the single estimator.
	
	\subsubsection{Deep Q-Network (DQN)}
	extends Q-learning with neural network to solve complex problems with extensive state spaces. It uses an online neural network parametrized by $\theta$ to approximate the vector of action values $Q(s,\cdot;\theta)$ for each state $s$, and a target network parameterized by $\theta'$ which is periodically copied from $\theta$ to reduce oscillation during training. The neural network is optimized by minimizing the difference between the predicted value $Q(s_t, a_t; \theta_{t})$ and the target value $Y^{Q}_{t} = r_{t+1} + \gamma\max_a Q(s_{t+1}, a; \theta_t^\prime)$, using experienced samples $(s_t, a_t, r_{t+1}, s_{t+1})$ drawn from a replay memory. To minimize the difference, the parameters of the network are updated along with the direction of the target value $Y_t^Q$ estimated by experienced samples $(s_t, a_t, r_{t+1}, s_{t+1})$ drawn from a replay memory using the following formula: 
	\begin{equation}
	\label{eq:gradient-update}
	\theta_{t+1} = \theta_{t} + \alpha \mathbb{E}[(Y_t^Q - Q(s_t, a_t; \theta_{t}))\nabla_{\theta_{t}}Q(s_t, a_t; \theta_{t})],
	\end{equation}
	where $ \nabla_{\theta_{t}}Q(s_t, a_t; \theta_{t}) $ is the gradient. Both the replay memory and the target network help DQN to stabilize learning and can dramatically improve its performance. However, like tabular Q-learning, using the single maximum estimator is prone to cause overestimating, leading to poor performance in many situations.
	
	\subsubsection{Double Q-learning} 
	uses the double estimator to ease the overestimation. The double estimator selects the action with the maximum Q value and evaluates the Q values of different actions separately in turn \cite{hasselt2010double}. The double Q-learning algorithm stores two Q-values, denoted $ Q^U $ and $ Q^V $, and replaces the estimated value $\max_{a^\prime}Q(s^\prime, a^\prime)$ in Equation \ref{eq:q-learning} with the combination $Q^U(s^\prime, \arg\max_{a^\prime}Q^V(s^\prime, a^\prime))$. Unfortunately, Hasselt  \cite{hasselt2010double} proved that though the double estimator can overcome the overestimation issue, a negative bias is introduced in the same time which may harm the resulting algorithm's performance and effectiveness.
	
	\subsubsection{Double DQN}
	incorporates the idea of double Q-learning into DQN to avoid the overestimation \cite{van2016deep}. It uses two sets of Q-networks $Q(s,a;\theta)$ and $Q(s,a,\theta^\prime)$: one for selecting action and the other for estimating the target Q-value. At each time the Q-network $Q(s,a;\theta)$ is updated using the following target value:
	\begin{equation}\label{eq5}
	Y_t^{Q} \equiv R_{t+1} + \gamma Q(s_{t+1}, \arg\max_a Q(s_{t+1}, a, \theta_t); \theta_t^\prime).
	\end{equation}
	By leveraging the above two Q-networks to select and evaluate the Q-values symmetrically in turn, this algorithm takes advantage of the double estimator to reduce the overestimation of Q values and lead to better performance in a variety of complex RL scenarios.
	
	\subsubsection{Weighted Double Q-learning (WDQ)}
	uses a dynamic heuristic value $ \beta $, which depends on a constant $ c $, to balance between the overestimation of the single estimator and the underestimation of the double estimator during the iterative Q-value update process:
	\begin{equation}\label{eq10}
	Q(s,a)^{U,WDQ} = \beta Q^U(s,a^*) + (1-\beta)Q^V(s,a^*),
	\end{equation}
	where a linear combination of $Q^U$ and $Q^V$ is used for updating Q-value. When $a^*$ is chosen by $Q^U, i.e., a^* \in \arg\max_aQ^U(s,a)$, $Q^U(s,a^*)$ will be positively biased and $Q^V(s,a^*)$ will be negatively biased, and vice versa. $\beta \in \left[ 0,1\right]$ balances between the positive and negative biases. Experiments on tabular MDP problems show that more accurate value estimation can indeed boost Q-learning's performance. However, it is still not clear whether this idea can be extended to the end-to-end DRL framework to handle high-dimensional problems.
	
	\subsubsection{Lenient Q-learning}
	\cite{potter1994cooperative} updates the policies of multiple agents towards an optimal joint policy simultaneously by letting each agent adopt an optimistic dispose at the initial exploration phase. This has been empirically proved to be efficient at increasing the likelihood of discovering the optimal joint policy in stochastic environments and avoiding agents gravitating towards a sub-optimal joint policy \cite{bloembergen2015evolutionary,palmer2017lenient,panait2006lenient,wei2016lenient}. 
	
	During training, lenient agents keep track of the temperature $T_t(s,a)$ for each state-action pair ($ s,a $) at time $ t $, which is initially set to a defined maximum temperature value and used for measuring the leniency $ l(s,a) $ as follows: 
	\begin{equation}
	l(s_t; a_t) = 1 - e^{-K * T_t(s_t, a_t)},
	\end{equation}
	where $ K $ is a constant determining how the temperature affects the decay in leniency. As suggested by \cite{wei2016lenient}, $T_t(s_t, a_t)$ is decayed using a discount factor $\kappa \in [0, 1]$ and $T_{t+1}(s_t, a_t)=\kappa T_t(s_t, a_t)$. Given the TD error $\delta = Y_t^Q - Q_t(s_t, a_t;\theta_t)$, the iterative update formula of lenient Q-learning is defined as follows:
	
		\begin{equation}
	\label{eq:lenient-q}
	Q(s_t, a_t) = \left\{ {\begin{array}{*{20}{l}}
		Q(s_t, a_t) + \alpha\delta & {\text{ if } \delta > 0 \text{ or } x > l(s_t, a_t),  }\\
		Q(s_t, a_t) &{\text{ otherwise.}}
		\end{array}} \right.
	\end{equation}
	The random variable $x \sim U(0,1)$ is used to ensure that a negative update $\delta$ is performed with a probability $1-l(s_t, a_t)$. Due to the
	initial state-action pairs being visited more often than the later ones, the temperature values for states close to the initial state can decay rapidly. One solution to address this is to fold the average temperature $\bar{T}(s^\prime) = \frac{1}{|A|} \sum_{a_i\in A}{T(s^\prime, a_i)}$ for next state $s^\prime$ into the temperature that is being decayed for $(s_t, a_t)$ \cite{wei2016lenient}, as below:
	\begin{equation}
	\label{eq:lenient-temperature}
	T_{t+1}(s_t, a_t) = \kappa*\left\{ {\begin{array}{*{20}{l}}
		T_t(s_t, a_t)  & {\text{\textup{ if} $s'$ \textup{is terminal,}}}\\
		(1-\eta)*T_t(s_t, a_t)+\eta \bar{T_t}(s') &{\text{ otherwise.}}
		\end{array}} \right.
	\end{equation}
	where $\eta$ is a constant controlling the extent that $\bar{T}(s^\prime)$ is folded in. We absorb this interesting notion of forgiveness into our lenient reward network to boost the convergence in cooperative Markov games which will be explained later.
	
	\section{Weighted Double Deep Q-Networks}
	In the section, we introduce a new multiagent DRL algorithm, weighted double deep Q-networks (WDDQN), with two auxiliary mechanisms, i.e., the lenient reward approximation and the scheduled replay strategy, to achieve efficient coordination in stochastic multiagent environments. In these environments, reward could be extremely stochastic due to the environments' inherent characteristics and the continuous change of the coexisting agents' behaviors.
	
	For the stochastic rewards caused by the environments, WDDQN uses the combination of the weighted double estimator and the reward approximator to reduce the estimation error. As for the non-stationary coexisting agents, we incorporate the leniency from lenient Q-learning \cite{palmer2017lenient,panait2006lenient} into the reward approximator to provide an optimistic estimation of the expected reward under each state-action pair $r(s,a)$. In addition, directly applying prioritized experience replay \cite{schaul_prioritized_2015} in multiagent DRL leads to poor performance, as stored transitions can become outdated because agents update their policies simultaneously. To address this, we propose a scheduled replay strategy to enhance the benefit of prioritization by adjusting the priority for transition sample dynamically. In the remainder of this section, we will describe these facets in details.
	
	\subsection{Network Architecture}
\begin{figure}[h]
		\centering
		\includegraphics[width=.8\linewidth]{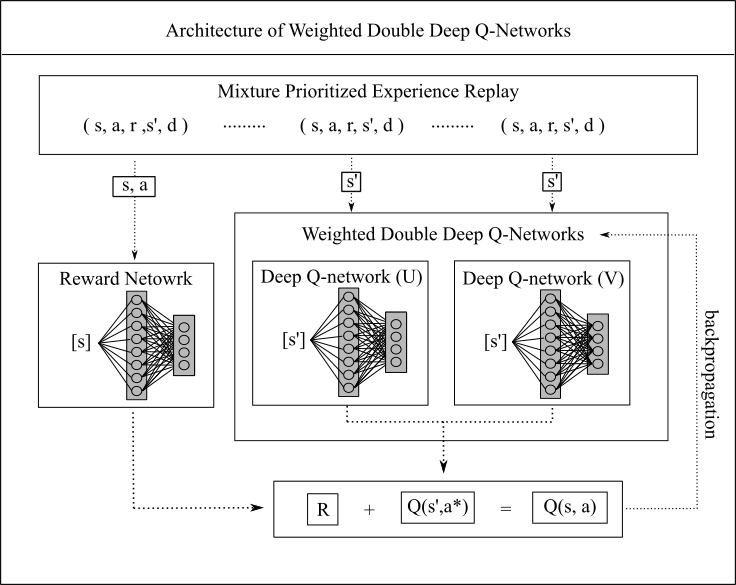}
		\caption{Network Architecture of WDDQN}
		\label{fig-reward-network}
	\end{figure}

	WDDQN outlined in Algorithm \ref{alg1} is adapted from WDQ by leveraging neural network as the Q-value approximator to handle problems with high-dimensional state spaces. The overall network architecture of the algorithm is depicted in Fig. \ref{fig-reward-network}. To reduce the estimation bias, WDDQN uses the combination of two estimators, represented as Deep Q-networks $Q^U$ and $Q^V$ with the same architecture, to select action $a = \max_{a'}\frac{Q^U(s,a') + Q^V(s,a')}{2}$ (line 5). Besides, the target $Q^{\tt Target}(s,a)$ (lines 12 and 17) used for Q-value updating in back-propagation is replaced with a weighted combination as well (lines 11 and 16). Intuitively, the combination balances between the overestimation and the underestimation. In addition, we also propose to use a reward approximator and an efficient scheduled replay strategy in WDDQN to achieve bias reduction and efficient coordination in multiagent stochastic environments.

	\begin{algorithm}[h]
		\caption{WDDQN}
		\label{alg1}
		\begin{algorithmic}[1]
			\State The maximum number of episodes: $Max_E$, the maximum number of steps: $Max_S$, global memory: $D^G$, episodic memory: $D^E$, reward network: $R^{N}$, deep Q-networks: $Q^U$ and $Q^V$
			\For{episode = 1 to $Max_E$}
			\State Initialize $D^E$
			\For{step = 1 to $Max_S$}
			\State $a \gets \max_{a'}\frac{Q^U(s,a') + Q^V(s,a')}{2}$ (with $\varepsilon$-greedy)
			\State Execute $a$ and store transitions into $D^E$
			
			\State Sample mini-batch $(s,a,r,s')$ of transitions from $D^G$
			\State Update $Q^U$ or $Q^V$ randomly
			\If {update $Q^U$}
			\State $a^* \gets \arg\max_aQ^U(s',a)$
			\State {$Q_U^w(s', a^*) \gets \beta Q^U(s',a^*) + (1-\beta) Q^V(s',a^*)$}
			\State $Q^{\tt Target}(s,a) \gets R^{N}(s,a) + Q_U^w(s', a^*)$
			\State{Update network $Q^U$ towards $Q^{\tt Target}$}
			\Else
			\State $a^* \gets \arg\max_aQ^V(s',a)$
			\State {$Q_V^w(s', a^*) \gets \beta Q^V(s',a^*) +(1-\beta) Q^U(s',a^*)$}
			\State $Q^{\tt Target}(s,a) \gets R^{N}(s,a) + Q_V^w(s', a^*)$
			\State{Update network $Q^V$ towards $Q^{\tt Target}$}
			\EndIf
			
			\State{Update $R^{N}$ according to transitions in $D^G$}
			
			\EndFor
			\State{Store $D^E$ into $D^G$}
			\EndFor
		\end{algorithmic}
	\end{algorithm}

	\subsection{Lenient Reward Network}
	To reduce noise in stochastic rewards, we use a reward network, which is a neural network estimator, to approximate the reward function $R(s,a) $ explicitly. The reward network can reduce bias in immediate reward $ r $ yielded from stochastic environments by averaging all rewards for distinct $(s,a)$ pair and be trained using the transitions stored in the experience replay during the online interaction. When updating the network, instead of using the reward $r$ in transition $(s,a,r,s^\prime)$ from experience memory, WDDQN uses the estimated reward by the reward network (lines 12 and 17).
	
	In addition to stochasticity, in a cooperative multiagent environment, the coexisting agents introduce additional bias to $ r $ as well. The mis-coordination of coexisting teammates may lower the reward $r$ for ($s$, $a^*$) despite the agent has adopted the optimal action. To address this, we use a lenient reward network (LRN) enhanced with the lenient concept in \cite{potter1994cooperative} to allow the reward network to be optimistic during the initial exploration phase. The LRN is updated periodically (line 20) as follows:
	
	\begin{equation}
	\label{eq:lenient-reward}
	R_{t+1}(s_t, a_t) =  \left\{{
		\begin{array}{*{20}{l}}
		R_t(s_t, a_t) + \alpha\delta	&{\textup{ if }\delta > 0 \text{ or } x < l(s_t, a_t),}		\\
		R_t(s_t, a_t)					&{\textup{ otherwise.}}
		\end{array}}\right.
	\end{equation}
	where $ R_t(s_t, a_t) $ is the reward approximation of state $ s $ and action $ a $ at time $ t $, and $\delta = \bar{r}_t^{(s,a)} - R(s_t,a_t)$ is the TD error between the $ R_t(s_t, a_t) $ and the target reward $ \bar{r}_t^{(s,a)} = 1/n \sum_{i = 1...n}{r_i^{(s,a)}} $ obtained by averaging all immediate reward $ r_i^{(s,a)} $ of $ (s,a) $ pairs in experience memory. Note that $l(s_t,a_t)$ inherits from Equation \ref{eq:lenient-q} and has the same meaning, which is gradually decayed each time a state-action $(s,a)$ pair is visited. Consequently, the LRN contributes to reduce bias by reward approximation and can help agents to find optimal joint policies in cooperative Markov games.
	
	\begin{figure}[h]
		\centering
		\includegraphics[width=\linewidth]{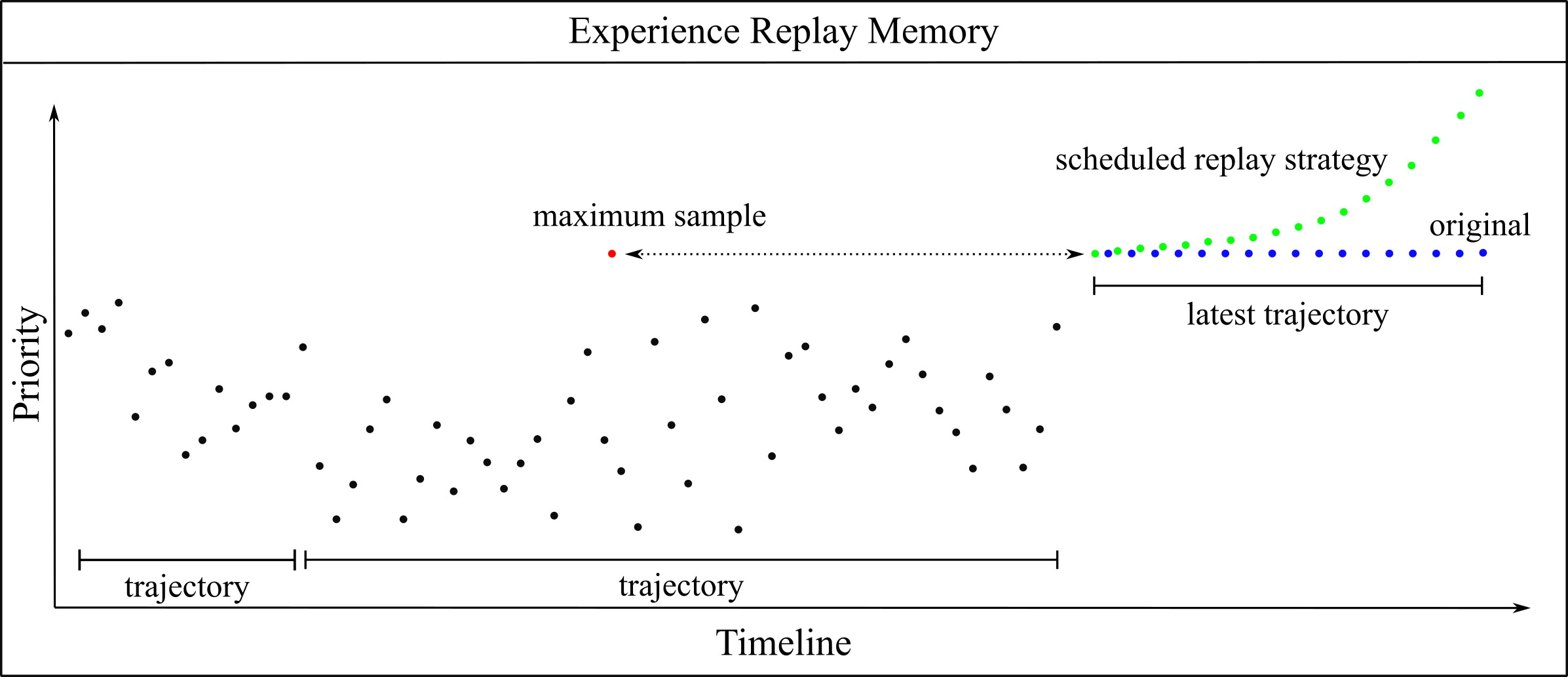}
		\caption{Comparison between the prioritized experience replay and the scheduled replay strategy: each dot represents a sample $(s,a,r,s)$, and a trajectory consists of an ordered sequence of samples. The x-axis represents the order that each sample comes into the relay memory and the y-axis is the priority of each sample. }
		\label{fig:variant}
	\end{figure}

	\begin{figure*}[t]
		\centering
		\includegraphics[width=\linewidth]{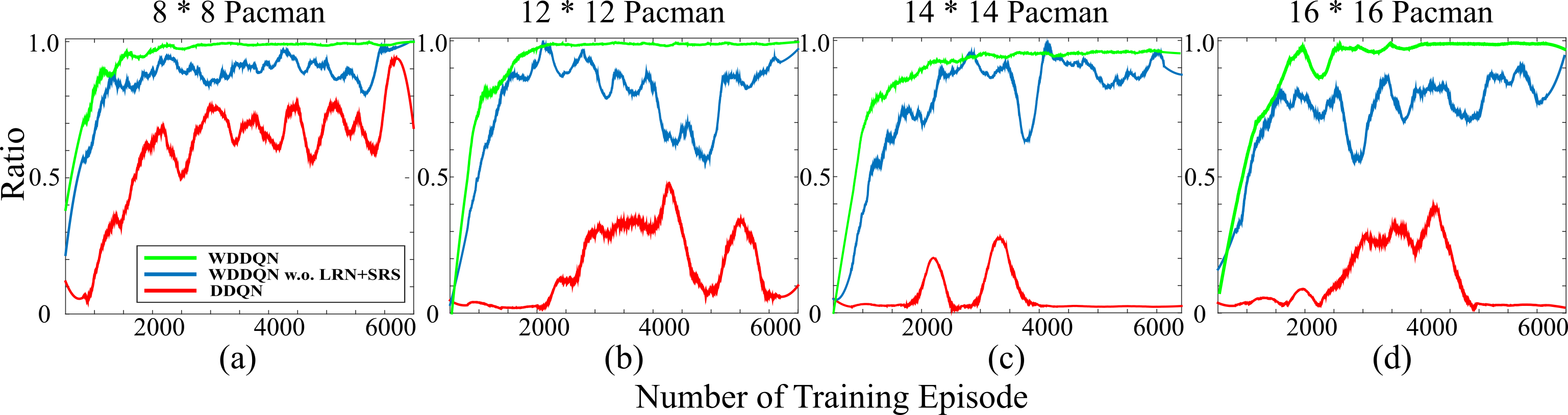}
		\caption{Comparisons of DDQN, WDDQN w.o. LRN+SRS and WDDQN on pacman with 4 different sizes. The X-axis is the number of training episodes and the Y-axis is a ratio of the number of minimum steps to the goal to the number of steps that the agent actually used during training.}
		\label{fig:pacman-training}
	\end{figure*}
	\subsection{Scheduled Replay Strategy}
	Prioritized experience replay (PER) can improve the DQN algorithm's training efficiency by allocating samples with different priorities according to their TD error. Samples with higher priorities are more likely to be chosen for network training. However, in stochastic multiagent environments, due to the noisy reward and the continuous behavior changes of coexisting agents, PER may deteriorate the algorithm's convergence and perform poorly. Given a transition $(s,a,r,s,d)$ with an extremely biased reward $ r $, PER will treat it as an important sample for its large TD error and will frequently select it to update the network, though it is incorrect due to the big noise in $ r $ at the beginning. To address this, we replace $ r $ with an estimation $ R^{N}(s,a)$ using LRN to correct TD error, by which the PER can distinguish truly important samples.
	
	Another potential problem is that PER gives all samples in the new trajectory the same priority, thus resulting in the indistinguishability of importance for all new samples. To be specific, in Fig. \ref{fig:variant}, the sample with the maximum priority is colored by red dot. PER gives all samples (blue dots) in the latest trajectory with an identical priority \footnote{See OpenAI source code for details: https://github.com/openai/baselines.}. However, in cooperative multiagent environments, the trajectories that agents succeed in cooperation are relatively rare, and in these trajectories the samples closer to the terminal state is even more valuable than the ones far from the terminal state. Besides, the $Q(s,a) = r + Q(s^\prime, a^*) $ far from the terminal state can further deteriorate if bootstrap of action value $Q(s^\prime, a^*)$ is already highly inaccurate, since inaccurate estimation will propagate throughout the whole contiguous samples. These two traits explain why samples that are close to the terminal state should be frequently used for network training. To this end, we develop a scheduled replay strategy (SRS) using a precomputed rising schedule $ [w_0,w_1,...,w_n] $ with size $ n $ to assign different priorities according to the sample's position $ i $ in the trajectory with $n$ samples.
	
	The values for $ w_i = e^{\rho_c*{u^i}} $ are computed using an exponent $ \rho^c $ which grows with a rising rate $ u > 1 $ for each $i$, $0 \leqslant i < n$. The priority $ p_i $ assigned to sample with index $ i $ is obtained by multiplying the current maximum priority $ p_{\max} $ in experience memory (priority of the red point in Fig. \ref{fig:variant}) by $ w_i $:
	\[
	p_i = p_{\max} \times w_i		
	\]
	
	The SRS assigns higher priority to samples near the terminal state (the green dot in Fig. \ref{fig:variant}) to ensure they are more likely to be sampled for network training. In this way, the estimation bias of the $ Q(s,a) $ near the terminal state is expected to decrease rapidly. This can significantly speed up the convergence and improve the training performance, as to be experimentally verified in the following section.
	
	\section{Experiments}
	Empirical evaluation is conducted to verify the effectiveness of WDDQN in terms of reducing bias and achieving coordination in stochastic multiagent domains. 
	
	First, we present comparisons of double DQN (DDQN) and WDDQN with /without LRN and SRS, denoted by WDDQN and WDDQN w.o. LRN+SRS, in terms of the bias reduction, learning speed and performance on a gridworld game with raw visual input. Then, we use a cooperative Markov game to investigate WDDQN's effectiveness of finding an optimal cooperative policy. A discussion about benefits of WDDQN, LNR and SRS is given in the end. 
	
	\begin{figure}[h]
		\centering
		\begin{minipage}[h]{.49\linewidth} 
			\centerline{\includegraphics[width =.5\textwidth]{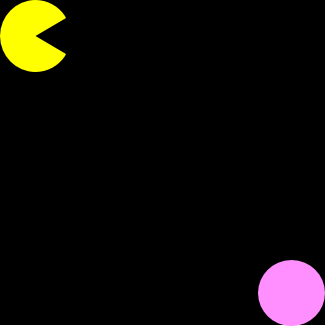}}
			\caption{Gridworld game.} \label{fig:picman}
		\end{minipage} 
		\hfill
		\begin{minipage}[h]{.49\linewidth}    
			\centerline{\includegraphics[width =.5\textwidth]{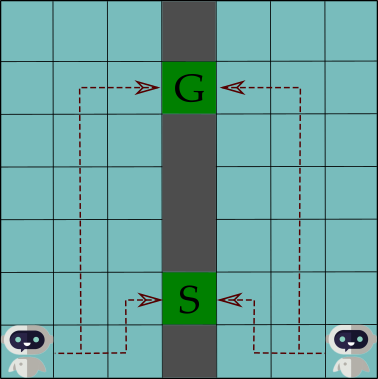}}
			\caption{Predator game.} \label{fig:predator}
		\end{minipage}
		\vfill
	\end{figure}
	\vspace{-0.5cm}
	\begin{table}[h]
		\caption{Network architectures in WDDQN}
		\label{tab:architecture}
		\centering
		\small
		\begin{tabular}{cccc}
			\hline 
			\# Network & Visual input & Filters in Conv. 1/2/3 & Unit in F.C \\
			\hline
			DQN & 84 * 84 * 3 & 32/64/64 & 512 \\
			LRN & 84 * 84 * 3 & 16/16/16 & 128\\
			\hline
		\end{tabular}
	\end{table}
	
	We set the constant $ c $ in $\beta$ to 0.1 in WDDQN, parameters $ K, \kappa, \eta $ in lenient Q-learning to 2, 0.95 and 0.6 respectively. Besides, the learning rate $\alpha$ for network training of DDQN, lenient Q-learning is set to 0.0001. Table \ref{tab:architecture} depicts the architecture of deep Q-networks and LRN in WDDQN. We use three hidden convolution layers (using rectifier non-linearities between each two consecutive layers), and a fully-connected hidden layer. The output layer of DQN and LRN is a fully-connect linear layer with a single output layer for each valid action $Q(s,a)$ and reward $R(s,a)$. For exploration purpose, DQN($ \epsilon $-greedy) is adopted with the $ \epsilon $ annealed linearly from 1 to 0.01 over the first 10000 steps. We used the Adam algorithm with 0.0001 learning rate and the minibatches of size 32. We trained for a total of 2500 episodes and used a replay memory of 8192 most recent frames. Last, to be fair, $ K, \kappa, \eta $ in LRN is the same as the lenient Q-learning while $\rho_c$ and $\mu$ in SRS is set to 0.2 and 1.1.
	
	\begin{figure*}[t]
		\centering
		\subfloat[deterministic rewards.]{\includegraphics[width=0.49\textwidth]{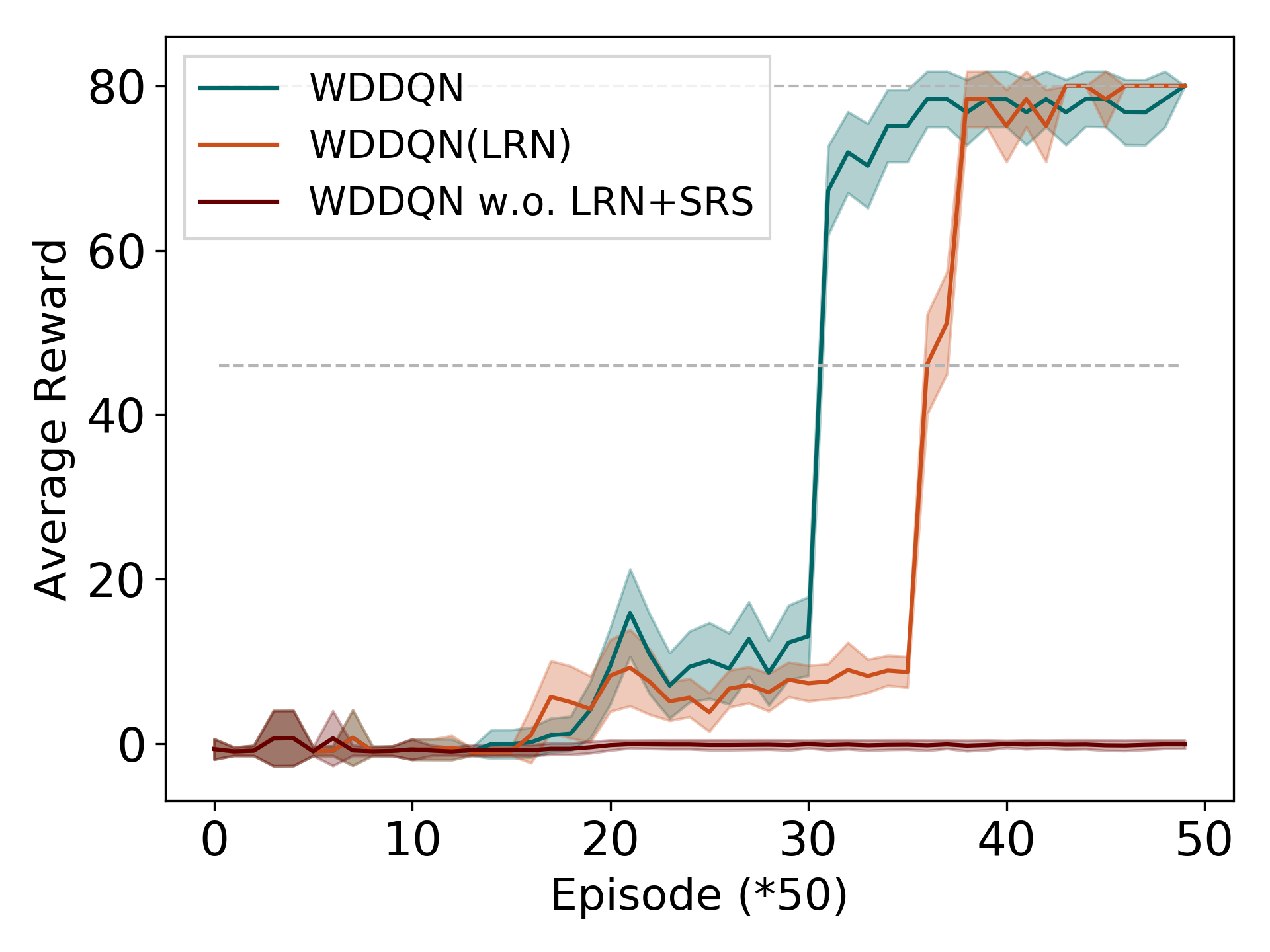}\label{fig:sf1}}
		\hfill
		\subfloat[stochastic rewards.]{\includegraphics[width=0.49\textwidth]{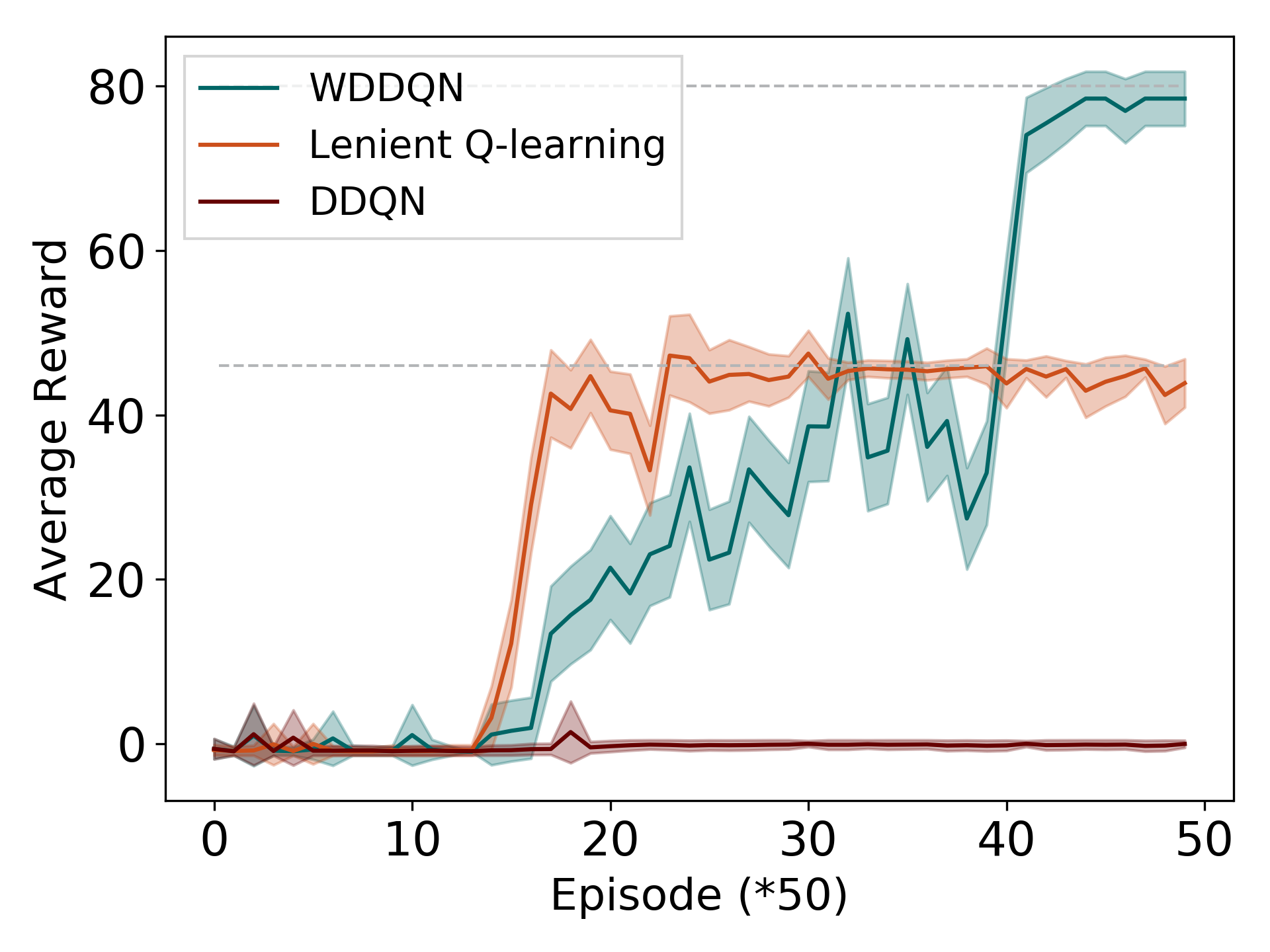}\label{fig:sf2}}
		\caption{(Left) Comparisons of WDDQN and its variants using the predator game with deterministic rewards; and (right) comparisons of WDDQN and other algorithms using the predator game with stochastic rewards. Note that, each point in the x-axis consists of 50 episodes, and the y-axis is the corresponding averaged reward.  The shadow area ranges from the lowest reward to the highest reward within the 50 episodes.}
		\label{fig:exp-pre}
	\end{figure*}
	\subsection{Pacman-like Grid World}
	The first experiment is an $n \times n$ pacman-like grid-world problem (Fig. \ref{fig:picman}), where the agent starts at the $s_0$ (top left cell), and moves towards the goal cell (pink dot at right bottom cell) using only four actions: \{north, south, east, west\}. Every movement leads the agent to move one cell in the corresponding direction, except that a collision on the edge of the grid results in no movement. The agent tries to search the goal cell which may appear randomly in any position in the grid world. The agent receives a stochastic reward of -30 or 40 with equal probability for any action entering into the goal and ending an episode. Choosing north or west will get a reward of -10 or +6, and south or east get a reward of -8 or +6 at a non-goal state. The environment is extremely noisy due to the uncertainty in the reward function.
		
	Empirical results in Figure \ref{fig:pacman-training} demonstrate that, under extremely stochastic environments, DDQN takes a long time to optimize the policy, while WDDQN w.o. LRN+SRS and WDDQN need much less episodes to get a better policy due to the weighted double estimator. DDQN and WDDQN w.o. LRN+SRS oscillate too frequently to converge to an optimal policy, while WDDQN performs steadily and smoothly because of the use of LRN. Another finding is that the training speed of WDDQN is faster than the others, which is attributed to the SRS. In general, WDQ works not as well as in relatively simple RL problems and both DDQN and WDDQN w.o. LRN+SRS may not converge even after a very long training time. By contrast, as shown in Fig. \ref{fig-reward-network}, WDDQN learns efficiently and steadily due to the use of both LRN and SRS.
	
	\subsection{Cooperative Markov Game}
	In this section, we consider the two predators pursuit problem. It is a more complex cooperative problem and firstly defined in \cite{benda1985optimal}. Here we redefine it in a simple way. The robots in Figure \ref{fig:predator} represent two agents trying to enter into the goal state at the same time. The cell with letter S is a suboptimal goal with a reward of +10 while G is a global optimal with a reward of +80. There is a thick wall (in gray) in the middle that separates the area into two zones. In each episode, two agents start at the left bottom cell and right bottom cell separately and try to go to the green goal cell together. Each agent has four actions: \{north, south, east, west\}. Every movement leads the agent to move one grid in the corresponding direction, except that a collision on the edge of the grid or thick wall results in no movement. A reward of 0 is received whenever entering into a non-goal state. The agent receives a positive reward for any action entering into the goal together and ending an episode, otherwise a negative reward of -1 is received with miscoordination.
	
	There are two types of cooperative policies moving towards the suboptimal goal cell S or the global optimal cell G, as shown in the Fig. \ref{fig:predator}. In the remaining part, we investigate whether WDDQN and related algorithms can find cooperative policies, especially the optimal policy.
	
	\subsubsection{Evaluation on WDDQN}
	Our goal is to train two agents simultaneously to coordinate in order to get higher rewards. The performance of WDDQN w.o. LRN+SRS, WDDQN(LRN)\footnote{WDDQN(LRN) uses only LRN and is identical to WDDQN w.o. SRS}, and WDDQN in terms of the average reward is depicted in Figure \ref{fig:exp-pre}\subref{fig:sf1}. As WDDQN w.o. LRN+SRS's convergence is no longer guaranteed in the neural network representation, it is not surprising that it fails in finding the cooperative policy by directly combining WDQ with neural network. By contrast, WDDQN(LRN), due to the LRN, achieves coordination more quickly and finds the optimal policy after a period of exploration. By leveraging the SRS, WDDQN shows a more promising result that the optimal policy is learned much faster than the two others.
	
	\subsubsection{Evaluation Against Other Algorithms}
	Here, we compare WDDQN against DDQN, a DRL algorithm, and lenient Q-learning, a multiagent RL algorithm on the same game except that the agent receives a reward of +10 or +100 with the possibility of 60\% or 40\% at goal S and a deterministic reward of +80 at goal G. Goal S is still suboptimal as its average reward is 46. This slight adjustment may affect the algorithm's performance by misleading the agent to converge to the suboptimal goal where a higher reward may appear accidentally. 
	
	Results in terms of the average reward are depicted in Fig. \ref{fig:exp-pre}\subref{fig:sf2}, where two dashed lines indicate optimal and suboptimal policy with the expected rewards of 80 and 46, respectively. Both WDDQN and lenient Q-learning outperform DDQN in terms of the convergence speed and the average reward in all experiments, which confirms the infeasibility of directly applying DRL algorithms in multiagent problems. Note that, WDDQN, due to the use of both LRN and SRS, is more stable, performs better and is more likely to find the optimal solution with the average reward of 80 than lenient Q-learning with the average reward of 46 in such a stochastic multiagent environment. 
	
	\section{Conclusion}
	This paper proposes WDDQN with the lenient reward network and the scheduled replay strategy to boost the training efficiency, stability and convergence under stochastic multiagent environments with raw image inputs, stochastic rewards, and large state spaces. Empirically, WDDQN performs better than WDDQN w.o. LRN+SRS, DDQN and lenient Q-learning in terms of the average reward and convergence rate on the pacman and two predators pursuit domains. 
	
	One downside to our approach is that it only uses one agent to explore the large-scale RL problems and train network at the same time. These can significantly slow down the exploration procedure and affect WDDQN's performance and efficiency. This could be remedied in practice by accelerating the training procedure of WDDQN using asynchronization, as being used in the A3C algorithm \cite{mnih2016asynchronous}. We leave this investigation to future work.
	
	\bibliographystyle{named}
	\bibliography{ijcai18}

\begin{thebibliography}{}

\bibitem[\protect\citeauthoryear{Bellman}{1957}]{bellman1958dynamic}
Richard Bellman.
\newblock {\em Dynamic programming}.
\newblock Princeton University Press., 1957.

\bibitem[\protect\citeauthoryear{Benda \bgroup \em et al.\egroup
  }{1986}]{benda1985optimal}
M.~Benda, V.~Jagannathan, and R.~Dodhiawala.
\newblock On optimal cooperation of knowledge sources - an empirical
  investigation.
\newblock Technical Report BCS--G2010--28, Boeing Advanced Technology Center,
  Boeing Computing Services, 1986.

\bibitem[\protect\citeauthoryear{Bloembergen \bgroup \em et al.\egroup
  }{2011}]{Bloembergen2011ETS}
Daan Bloembergen, Michael Kaisers, and Karl Tuyls.
\newblock Empirical and theoretical support for lenient learning.
\newblock In {\em International Conference on Autonomous Agents and Multiagent
  Systems}, pages 1105--1106, 2011.

\bibitem[\protect\citeauthoryear{Bloembergen \bgroup \em et al.\egroup
  }{2015}]{bloembergen2015evolutionary}
Daan Bloembergen, Karl Tuyls, Daniel Hennes, and Michael Kaisers.
\newblock Evolutionary dynamics of multi-agent learning: A survey.
\newblock {\em Journal of Artificial Intelligence Research}, 53:659--697, 2015.

\bibitem[\protect\citeauthoryear{Claus and Boutilier}{1998}]{claus1998dynamics}
Caroline Claus and Craig Boutilier.
\newblock The dynamics of reinforcement learning in cooperative multiagent
  systems.
\newblock In {\em AAAI Conference on Artificial Intelligence}, pages 746--752,
  1998.

\bibitem[\protect\citeauthoryear{Gupta \bgroup \em et al.\egroup
  }{2017}]{gupta2017cooperative}
Jayesh~K Gupta, Maxim Egorov, and Mykel Kochenderfer.
\newblock Cooperative multi-agent control using deep reinforcement learning.
\newblock In {\em International Conference on Autonomous Agents and Multiagent
  Systems}, pages 66--83, 2017.

\bibitem[\protect\citeauthoryear{Hasselt}{2010}]{hasselt2010double}
Hado~V. Hasselt.
\newblock Double {Q}-learning.
\newblock In {\em Advances in Neural Information Processing Systems}, pages
  2613--2621, 2010.

\bibitem[\protect\citeauthoryear{Lanctot \bgroup \em et al.\egroup
  }{2017}]{lanctot2017unified}
Marc Lanctot, Vinicius Zambaldi, Audrunas Gruslys, Angeliki Lazaridou, Julien
  Perolat, David Silver, Thore Graepel, et~al.
\newblock A unified game-theoretic approach to multiagent reinforcement
  learning.
\newblock In {\em Advances in Neural Information Processing Systems}, pages
  4193--4206, 2017.

\bibitem[\protect\citeauthoryear{Matignon \bgroup \em et al.\egroup
  }{2007}]{matignon2007}
La{\"e}titia Matignon, Guillaume~J Laurent, and Nadine Le~Fort-Piat.
\newblock Hysteretic q-learning: an algorithm for decentralized reinforcement
  learning in cooperative multi-agent teams.
\newblock In {\em International Conference on Intelligent Robots and Systems},
  pages 64--69, 2007.

\bibitem[\protect\citeauthoryear{Matignon \bgroup \em et al.\egroup
  }{2012}]{matignon2012independent}
Laetitia Matignon, Guillaume~J Laurent, and Nadine Le~Fort-Piat.
\newblock Independent reinforcement learners in cooperative markov games: a
  survey regarding coordination problems.
\newblock {\em The Knowledge Engineering Review}, 27(1):1--31, 2012.

\bibitem[\protect\citeauthoryear{Mnih \bgroup \em et al.\egroup
  }{2013}]{mnih_playing_2013}
Volodymyr Mnih, Koray Kavukcuoglu, David Silver, Alex Graves, Ioannis
  Antonoglou, Daan Wierstra, and Martin Riedmiller.
\newblock Playing atari with deep reinforcement learning.
\newblock {\em arXiv preprint arXiv:1312.5602}, 2013.

\bibitem[\protect\citeauthoryear{Mnih \bgroup \em et al.\egroup
  }{2015}]{mnih_human-level_2015}
Volodymyr Mnih, Koray Kavukcuoglu, David Silver, Andrei~A. Rusu, Joel Veness,
  Marc~G. Bellemare, Alex Graves, Martin Riedmiller, Andreas~K. Fidjeland,
  Georg Ostrovski, Stig Petersen, Charles Beattie, Amir Sadik, Ioannis
  Antonoglou, Helen King, Dharshan Kumaran, Daan Wierstra, Shane Legg, and
  Demis Hassabis.
\newblock Human-level control through deep reinforcement learning.
\newblock {\em Nature}, 518(7540):529--533, 2015.

\bibitem[\protect\citeauthoryear{Mnih \bgroup \em et al.\egroup
  }{2016}]{mnih2016asynchronous}
Volodymyr Mnih, Adria~Puigdomenech Badia, Mehdi Mirza, Alex Graves, Timothy
  Lillicrap, Tim Harley, David Silver, and Koray Kavukcuoglu.
\newblock Asynchronous methods for deep reinforcement learning.
\newblock In {\em International Conference on Machine Learning}, pages
  1928--1937, 2016.

\bibitem[\protect\citeauthoryear{Palmer \bgroup \em et al.\egroup
  }{2018}]{palmer2017lenient}
Gregory Palmer, Karl Tuyls, Daan Bloembergen, and Rahul Savani.
\newblock Lenient multi-agent deep reinforcement learning.
\newblock In {\em International Conference on Autonomous Agents and Multi-Agent
  Systems}, page to appear, 2018.

\bibitem[\protect\citeauthoryear{Panait \bgroup \em et al.\egroup
  }{2006}]{panait2006lenient}
Liviu Panait, Keith Sullivan, and Sean Luke.
\newblock Lenient learners in cooperative multiagent systems.
\newblock In {\em International Conference on Autonomous Agents and Multiagent
  Systems}, 2006.

\bibitem[\protect\citeauthoryear{Potter and
  De~Jong}{1994}]{potter1994cooperative}
Mitchell~A Potter and Kenneth~A De~Jong.
\newblock A cooperative coevolutionary approach to function optimization.
\newblock In {\em International Conference on Parallel Problem Solving from
  Nature}, pages 249--257, 1994.

\bibitem[\protect\citeauthoryear{Schaul \bgroup \em et al.\egroup
  }{2016}]{schaul_prioritized_2015}
Tom Schaul, John Quan, Ioannis Antonoglou, and David Silver.
\newblock Prioritized experience replay.
\newblock In {\em International Conference on Learning Representations}, 2016.

\bibitem[\protect\citeauthoryear{Smith and Winkler}{2006}]{smith2006optimizer}
James~E Smith and Robert~L Winkler.
\newblock The optimizer’s curse: Skepticism and postdecision surprise in
  decision analysis.
\newblock {\em Management Science}, 52(3):311--322, 2006.

\bibitem[\protect\citeauthoryear{Sutton and
  Barto}{1998}]{sutton1998reinforcement}
Richard~S Sutton and Andrew~G Barto.
\newblock {\em Reinforcement learning: An introduction}.
\newblock MIT press Cambridge, 1998.

\bibitem[\protect\citeauthoryear{Sutton}{1988}]{sutton1988learning}
Richard~S Sutton.
\newblock Learning to predict by the methods of temporal differences.
\newblock {\em Machine learning}, 3(1):9--44, 1988.

\bibitem[\protect\citeauthoryear{Van~Hasselt \bgroup \em et al.\egroup
  }{2016}]{van2016deep}
Hado Van~Hasselt, Arthur Guez, and David Silver.
\newblock Deep reinforcement learning with double q-learning.
\newblock In {\em AAAI Conference on Artificial Intelligence}, pages
  2094--2100, 2016.

\bibitem[\protect\citeauthoryear{Wang \bgroup \em et al.\egroup
  }{2016}]{wang2015dueling}
Ziyu Wang, Tom Schaul, Matteo Hessel, Hado Van~Hasselt, Marc Lanctot, and Nando
  De~Freitas.
\newblock Dueling network architectures for deep reinforcement learning.
\newblock In {\em International Conference on Learning Representations}, 2016.

\bibitem[\protect\citeauthoryear{Watkins}{1989}]{watkins1989learning}
Christopher John Cornish~Hellaby Watkins.
\newblock {\em Learning from delayed rewards}.
\newblock PhD thesis, King's College, University of Cambridge, 1989.

\bibitem[\protect\citeauthoryear{Wei and Luke}{2016}]{wei2016lenient}
Ermo Wei and Sean Luke.
\newblock Lenient learning in independent-learner stochastic cooperative games.
\newblock {\em Journal of Machine Learning Research}, 17(84):1--42, 2016.

\bibitem[\protect\citeauthoryear{Zhang \bgroup \em et al.\egroup
  }{2017}]{zhangweighted}
Zongzhang Zhang, Zhiyuan Pan, and Mykel~J Kochenderfer.
\newblock Weighted double {Q}-learning.
\newblock In {\em International Joint Conference on Artificial Intelligence},
  pages 3455--3461, 2017.

\end{thebibliography}
\end{document}